\documentclass[runningheads]{llncs}
\usepackage{graphicx}
\usepackage{comment}
\usepackage{amsmath,amssymb} 
\usepackage{color}

\usepackage{subcaption}
\usepackage{epsfig}
\usepackage{graphicx}
\usepackage{amsmath}
\usepackage{amssymb}
\usepackage{bm}
\usepackage{adjustbox}
\usepackage{slashbox}
\usepackage{float}
\usepackage{booktabs}
\usepackage{multirow}
\usepackage{caption}

\usepackage{xspace}
\usepackage{cleveref}
\usepackage{tabularx}

\makeatletter
\DeclareRobustCommand\onedot{\futurelet\@let@token\@onedot} 
\def\@onedot{\ifx\@let@token.\else.\null\fi\xspace}

\def\eg{\emph{e.g}\onedot} 
\def\ie{\emph{i.e}\onedot} 
\def\cf{\emph{c.f}\onedot}

\def\etal{\emph{et al}\onedot}
\makeatother

\begin{document}
\pagestyle{headings}
\mainmatter
\def\ECCVSubNumber{1809}  
\setlength{\parskip}{0cm}


\title{Low Light Video Enhancement using Synthetic Data Produced with an Intermediate Domain Mapping} 


\titlerunning{Low Light Video Enhancement using Synthetic Data}
%
\author{
Danai Triantafyllidou \and
Sean Moran  \and
Steven McDonagh \and 
Sarah Parisot \and 
Gregory Slabaugh }

\authorrunning{D. Triantafyllidou, S. Moran, et al.}
%
\institute{Huawei Noah’s Ark Lab  \\
\vspace{2mm}
\email{\{danaitri22, sean.j.moran\}@gmail.com} \\
\email{\{steven.mcdonagh, sarah.parisot, gregory.slabaugh\}@huawei.com}}
\maketitle

\begin{abstract}
Advances in low-light video RAW-to-RGB translation are opening up the possibility of fast low-light imaging on commodity devices (\eg smartphone cameras) without the need for a tripod. However, it is challenging to collect the required paired short-long exposure frames to learn a supervised mapping.  Current approaches require a specialised rig or the use of \emph{static} videos with no subject or object motion, resulting in datasets that are limited in size, diversity, and motion.  We address the data collection bottleneck for low-light video RAW-to-RGB by proposing a data synthesis mechanism, dubbed \emph{SIDGAN}, that can generate abundant dynamic video training pairs. SIDGAN maps videos found `in the wild' (\eg{}~internet videos) into a low-light (short, long exposure) domain. By generating dynamic video data synthetically, we enable a recently proposed state-of-the-art RAW-to-RGB model to attain higher image quality (improved colour, reduced artifacts) and improved temporal consistency, compared to the same model trained with only static real video data.
\end{abstract}

\section{Introduction}
\label{sec:intro}

Low-light imaging (less than 5 lux) is a challenging scenario for camera image signal processor (ISP) pipelines due to the low photon count, low signal-to-noise ratio (SNR) and profound colour distortion~\cite{Chen2018learning}. The ISP is responsible for forming a high-quality RGB image with minimal noise, pleasing colors, sharp detail, and good contrast from the originally captured RAW data. Recently there has been growing research interest in end-to-end deep neural network architectures for modelling the entire ISP pipeline, both in well-lit~\cite{schwartz19} and low-light scenarios~\cite{Chen2018learning}.

\begin{figure*}
\begin{center}
\begin{tabular}{c@{}c@{}c@{}}
    \scalebox{1}{Ground Truth} &
    \scalebox{1}{SID motion} &
    \scalebox{1}{Ours (SIDGAN)} \\
     \includegraphics[scale=0.08]{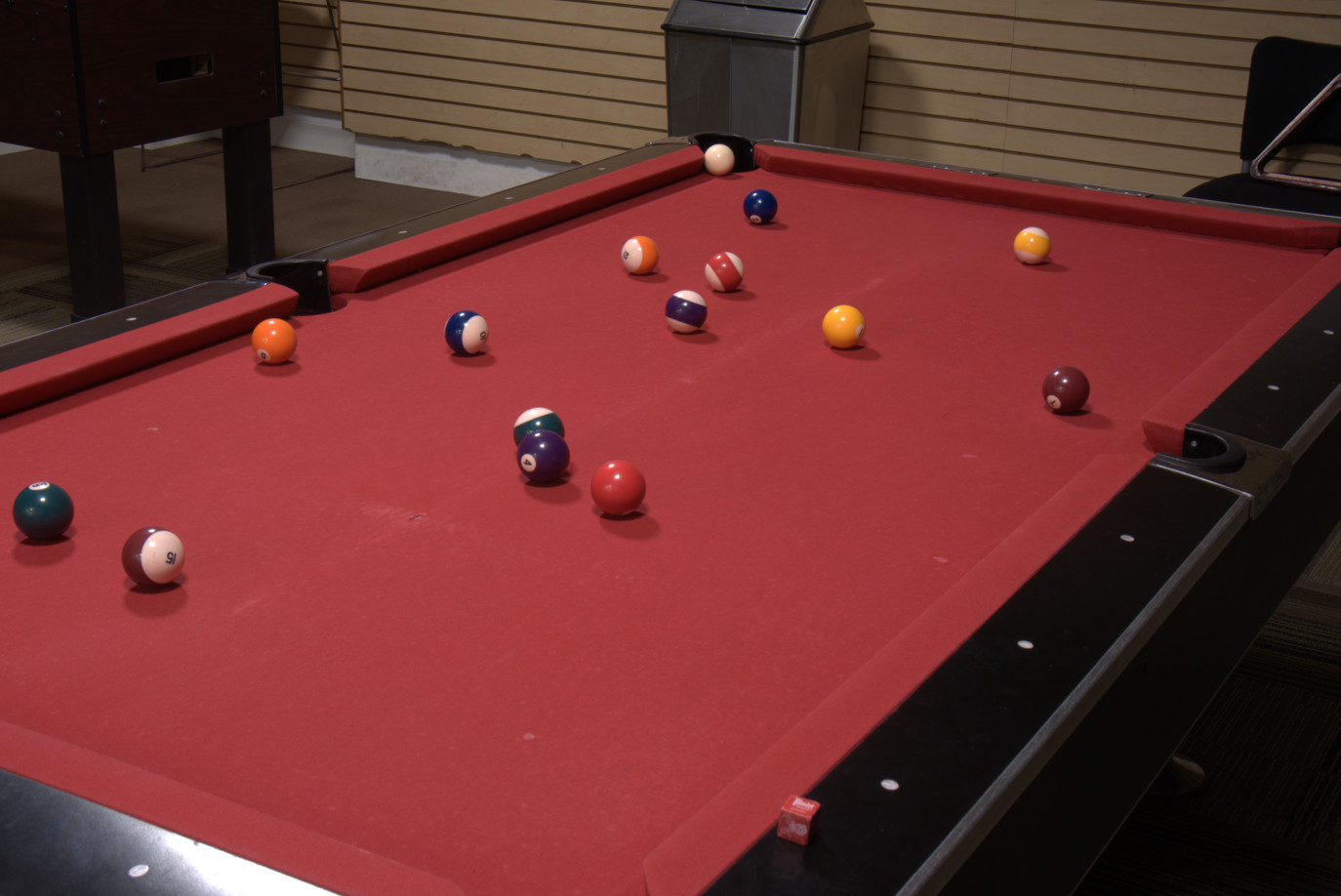} &
\includegraphics[scale=0.08]{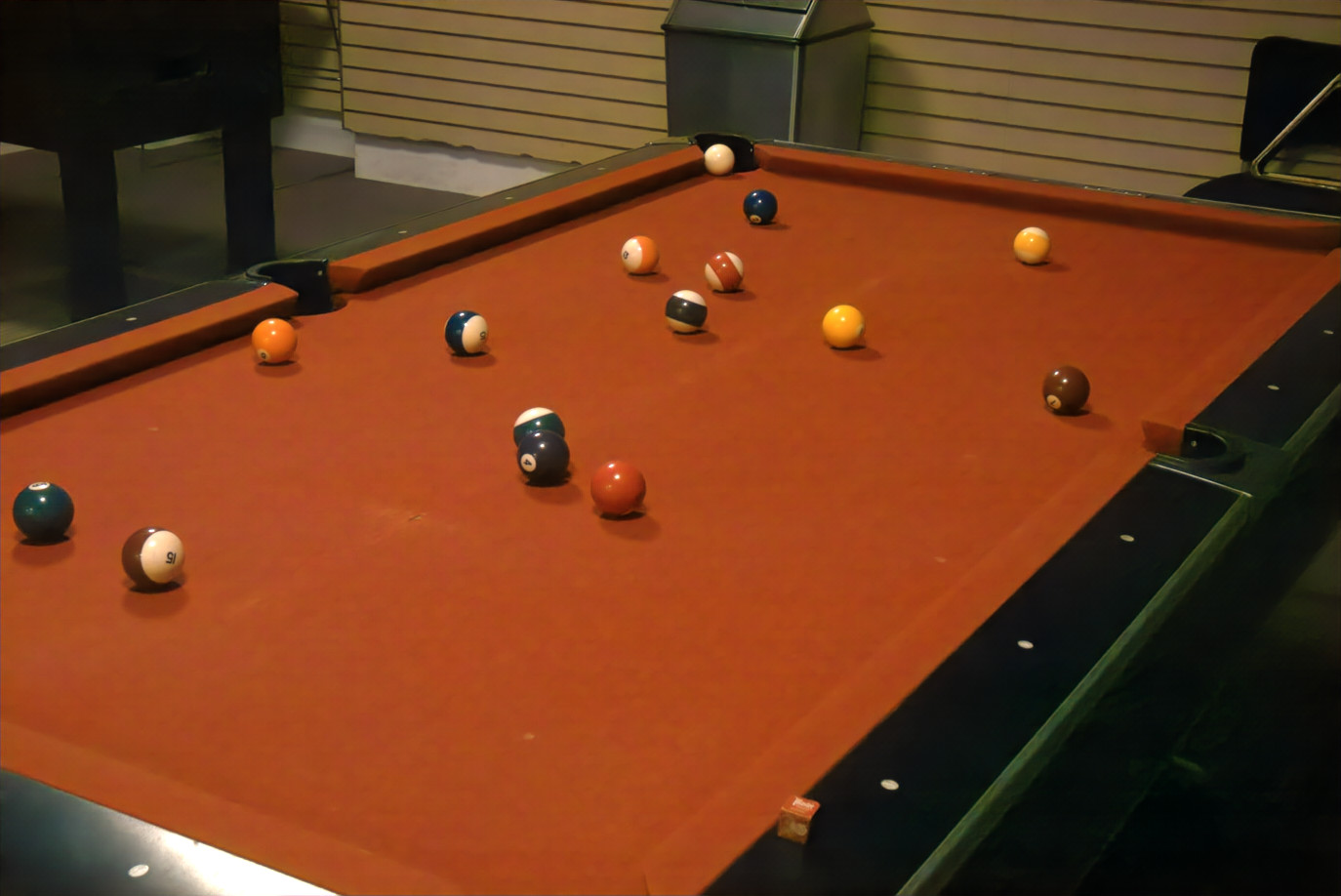} &
\includegraphics[scale=0.08]{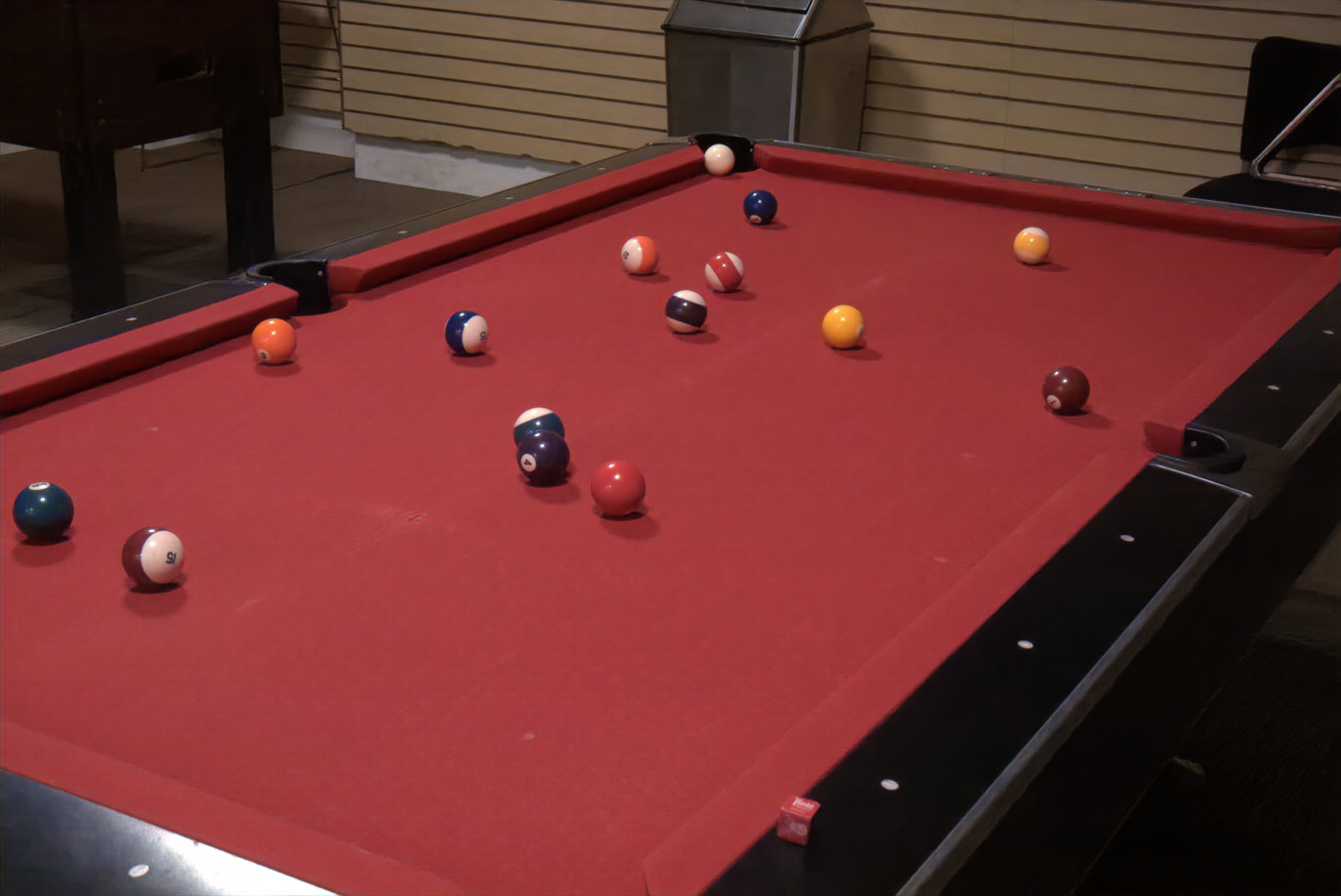} \\
\end{tabular}
\end{center}
\caption{Comparing the image quality with SID motion \cite{Jiang_2019_ICCV}. Training with synthetic data provides a more capable colour reproduction}
\label{fig:visual_comparison1}
\end{figure*}

A major bottleneck in the learning of deep models for end-to-end RAW-to-RGB mapping is the availability of data. Existing models require a large amount of manually collected paired data (RAW sensor data and its corresponding RGB image) for training. However, collecting suitable amounts of paired data is often time consuming, error prone (\eg misaligned pairs), and expensive. Chen \etal~\cite{Chen_2019_ICCV} resort to using a tripod to collect static videos for training. In~\cite{Jiang_2019_ICCV} a novel optical system is designed to obtain dark and bright paired frames of the same scene simultaneously, but this rig is not publicly available, requires expertise to operate and only works if the scene is adequately illuminated. These challenges result in datasets that are limited in size, diversity in scene type, content and motion. This in turn typically produces models offering only limited colour reproduction and temporal consistency on real dynamic video (Figure~\ref{fig:visual_comparison1} and Section~\ref{sec:experiments}). 
In this paper, we address the training data bottleneck for learning the RAW-to-RGB mapping, with a specific focus on low-light dynamic synthetic video data generation. Low-light video enhancement provides an ideal testbed for studying the potential of synthetic data as it is highly challenging to manually collect such data. In contrast to pre-existing work, we propose \textbf{S}eeing \textbf{I}n the \textbf{D}ark \textbf{GAN} (SIDGAN), a synthetic low-light video data generation pipeline leveraging Generative Adversarial Networks (GANs)~\cite{Goodfellow2014}.

GANs have proved to be a powerful modelling paradigm for learning complex high dimensional data manifolds for many types of real-world data such as natural images. The data distribution is modelled by framing learning as a competition between a generator network and a discriminator network. The generator attempts to produce samples from the desired data distribution that are as realistic as possible such that the discriminator network is fooled into classifying the synthetic samples as being real. The ensuing minimax game between the two networks can lead to a generator network that produces realistic samples from the data manifold. SIDGAN builds on the CycleGAN work of Zhu \etal~\cite{zhu2017unpaired} who demonstrate how to learn an unpaired mapping between two disparate domains (\eg two sets of images with different styles). However, different to their work, we extend the mapping to three domains using a pair of CycleGANs while leveraging a weak supervisory signal in the form of an \emph{intermediate} domain that has a paired data relationship with one of the remaining two domains. We argue that for an effective mapping between two domains that are very distant (\eg internet videos and short exposure frames from a completely different sensor), that it is best to leverage an intermediate domain. Our approach is illustrated in Figure~\ref{fig:pipeline}.

Our main contributions are three-fold:

\begin{itemize}
 
     \item \textbf{Semi-supervised dual CycleGAN with intermediate domain mapping}: Mapping directly from internet videos (Figure~\ref{fig:pipeline}, domain A) to short exposure (domain C) is difficult due to the large domain gap and lack of paired training examples. Instead, we bridge the gap using an intermediate long exposure domain (domain B) for which we have paired data (between domains B and C).  This decomposes a difficult problem into two simpler problems, the latter with supervision. 
     \item \textbf{Data abundance for RAW-to-RGB video}: The Dual CycleGAN allows synthesis of abundant video data in order to train high capacity models with, typically unavailable, dynamic and domain specific paired training data.
     \item \textbf{A practical strategy to combine synthetic and real data}:  We propose an effective three-step training and fine-tuning scheme to address the remaining domain gap between synthetically generated and real video data. Combining our dynamic synthetic data with static real data yields a forward RAW-to-RGB video model with superior temporal consistency and colour reproduction compared to the same model trained with only real data.
\end{itemize}

\section{Related Work}
\label{sec:related_work}

Low-light image and video quality enhancement topics are closely related to our contributions. In addition to these areas, we briefly review intermediate domain mappings, synthetic image generation and learning with unpaired data.

\noindent \textbf{Low-light image enhancement.} A large body of work exists on low-light image enhancement, spanning histogram equalization (HE) techniques~\cite{ibrahim2007brightness,arici2009histogram,nakai2013color} and approaches grounded in Retinex theory~\cite{land1977retinex,jobson1997multiscale,jobson1997properties,ying2017new}. Classical enhancement methods often make use of statistical techniques that typically rely on strong modeling assumptions, which may not hold true in real world scenes or scenarios. Deep learning techniques have also been readily applied to low-light image enhancement in recent years. The work of LLNet~\cite{lore2017llnet} employed an autoencoder towards low-light image denoising. Further convolutional works have used multiscale feature maps~\cite{tao2017llcnn} and brightness transmission image priors~\cite{tao2017low} to enhance image contrast with strong qualitative results \cf classical approaches.

\noindent \textbf{Low-light video enhancement.} The video enhancement problem is more recent and has received comparatively less attention. Analogous to static images, statistical Retinex theory has also been applied to video~\cite{liu2016low,wang2014piecewise}. Framing the problem in a joint-task setting was investigated by~\cite{kim2015novel}; coupling low-light enhancement with denoising.
Network based learning is also considered for video; Lv \etal~\cite{Lv2018MBLLEN} propose a multi-branch low-light enhancement network, applicable to both image and video domains. As earlier highlighted, learning-based mapping of (low-light) RAW-to-RGB work is highly relevant for our direction; Chen \etal~\cite{Chen2018learning,Chen_2019_ICCV} learn this transformation considering both images and, latterly, video.

Capture of real video data in this problem setting is prohibitively expensive. However, as noted, systems have been proposed that can capture both bright and dark videos of identical scene content, providing training pairs for low-light video models.  
Jiang \etal~\cite{Jiang_2019_ICCV} collected data and employed a standard CNN to learn enhancement mappings for the transformation from low-light raw camera sensor data to bright RGB videos. Collected data was relatively small by deep learning standards ($179$ video pairs), illustrating the arduous burden of real-world video collection in scenarios that involve complex capture setups and custom hardware (here beam splitters and relay lenses). Uncommon specialist hardware, operator expertise requirements and support for (only) adequately illuminated scenes can be considered the main disadvantages of video enhancement work that depends exclusively on real-world data.
 
\noindent \textbf {Intermediate domain mappings.} The concept of harnessing intermediate domain bridges can be considered powerful and related strategies have been employed in a number of scenarios~\cite{Gong_2019_CVPR,tan2017distant,li2019learning,gopalan2011domain,cui2014flowing}. In addition to visual domains, evidence in support of the broader applicability of this family of strategies is also found in machine translation tasks~\cite{DBLP:journals/corr/abs-1905-06831}, where intermediate domains enabled extension of bilingual systems to become multilingual. Relevant synthetic data work~\cite{han2019unsupervised}, that we draw from, leverages chains of image mappings (``indirect-paths'') to gain a supervisory signal towards improving super-resolution.
Combining intermediate domain mappings with synthetic data offers a promising direction for problem domains where acquisition of paired imagery, and therefore direct supervisory signal, is challenging.

\noindent \textbf{Synthetic data augmentation.} The use of synthetic data for model training and testing can be considered popular and datasets have been created for a multitude of image processing and computer vision problems~\cite{dosovitskiy2015flownet,mcdonagh2016synthetic,richter2016playing,gaidon2016virtual}. Early work performed successful scene text recognition with simplistic data generation~\cite{Jaderberg14c} and, more recently, the benefits of combining synthetic data with Generative Adversarial Networks (GANs)~\cite{Goodfellow2014} have been actively explored.

The work of~\cite{zhu2018data} explores GAN data augmentation, generating artificial images using conditional Generative Adversarial Networks (cGANs). By conditioning on segmentation masks, realistic images were generated for their task (leaf segmentation), and related performance improved by ${\sim}16\%$ \cf without synthetic augmentation. In~\cite{gecer2018semi} a semi-supervised adversarial framework is used to generate photorealistic facial images. By introducing pairwise adversarial supervision, two-way domain adaptation is constrained using only small paired real (and synthetic) images along with a large volume of unpaired data. Performance improves, due to the synthetic imagery, and consistently betters that of a face recognition network trained with Oxford VGG Face data. In~\cite{tripathy2018learning} both paired and unpaired training data is utilised simultaneously in conjunction with generative models. Two generators and four discriminators are employed in a hybrid setting and qualitatively strong results, on multiple image-to-image translation tasks, are reported. Mixed and fully unsupervised approaches~\cite{lugmayr2019unsupervised} begin to show great promise in faithfully generalising to real-world image distributions that are naturally sample-scarce or where data is otherwise hard to collect.  
These results motivate the use of inexpensive synthetic data for training GAN based tools. The need to collect large amounts of hand-annotated real-world data is avoided yet performance can surpass that of training with real-data exclusively. 

In contrast to these successes, recent work~\cite{ravuri2019seeing} reports interesting findings when employing  
generative models (\eg BigGAN), for data augmentation. 
Image classification error performance (Top-1 and Top-5)  
improves only marginally when additional synthetic data is added to an ImageNet training set.  
As GAN tools begin to be employed to aid downstream tasks, metrics that appropriately measure \emph{downstream task performance} must be utilised \cf solely evaluating synthetic sample image quality. 
Towards this our current work considers quantitative downstream performance evaluation, providing evidence towards the efficacy of our proposed data generation strategy (Section~\ref{sec:experiments}).

\section{Learning the Low-Light Video RAW-to-RGB Mapping}
\label{sec:methods}

Our objective is to learn short-to-long exposure mappings that provide accurate colour reproduction and temporal consistency. Given the lack of available real short, long exposure video pairs we propose a two-step approach that leverages \emph{data synthesis}. The first step (Section~\ref{sec:cyclegan}), involves training a dual CycleGAN model for the purposes of data synthesis. The dual CycleGAN maps video frames to a domain characterised by short exposure images. A domain bridge (\eg long exposure images) is used to regularise the mapping with available paired supervision. The trained CycleGAN permits videos `from the wild' to be projected into the long and then short exposure domains, thereby generating the necessary paired supervision. Our second step, detailed in Section~\ref{sec:forward}, utilises this synthetic data to train a forward model capable of mapping low-light video RAW to long exposure RGB. Finally, Section~\ref{sec:arch} provides details on synthetic data generation network architectures.

\subsection{Synthetic Data Generation Using an Intermediate Domain}\label{sec:cyclegan}
\begin{figure*}[t!]
  \centering
  \includegraphics[scale=0.42]{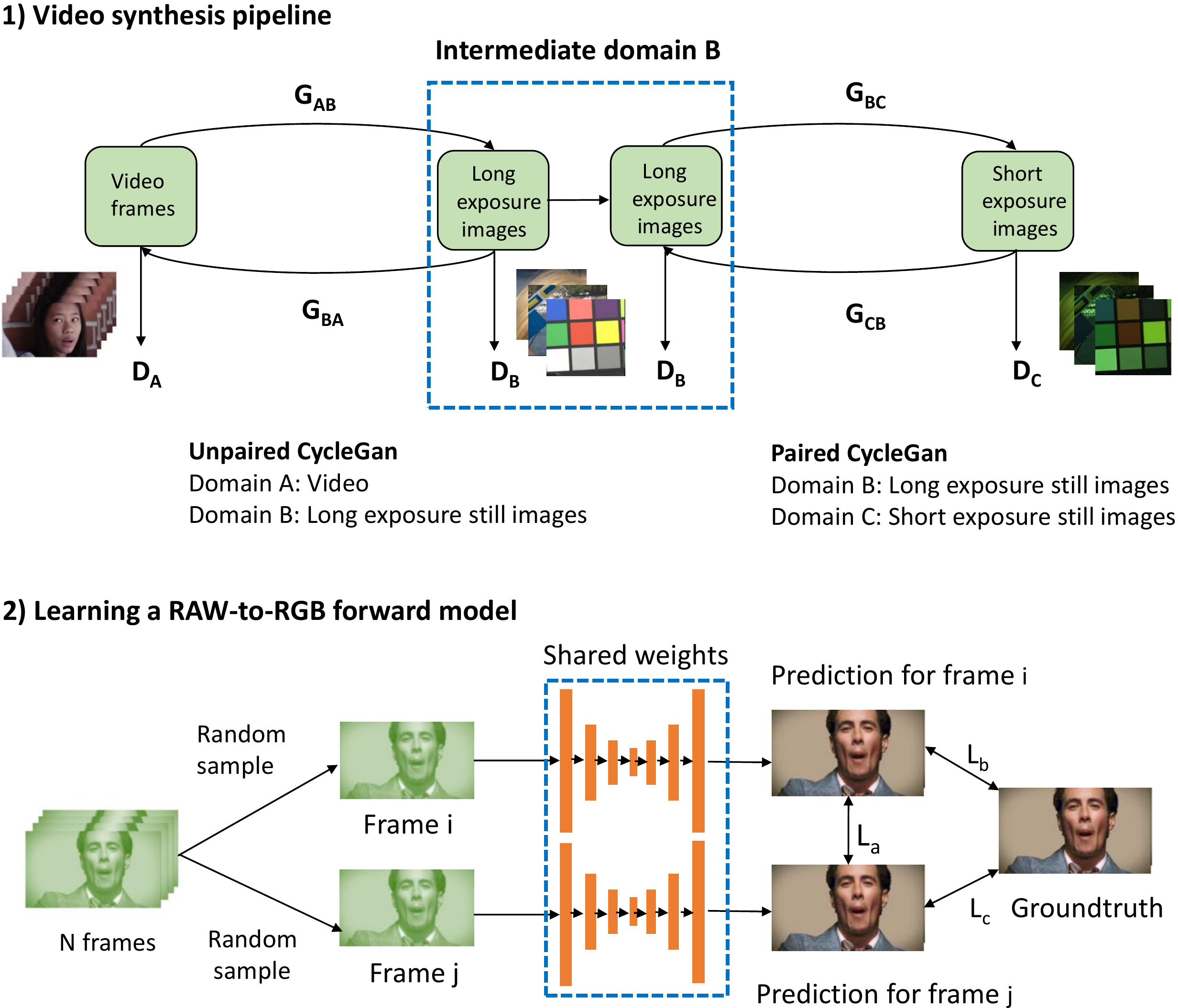}
  \caption{\textbf{Step 1:} We use SIDGAN generators ($G_{AB}$, $G_{BC}$) to map Vimeo videos (domain A) into the long (domain B), then short (domain C) exposure domains, giving our synthetic training dataset. \textbf{Step 2:} The forward model can be very different from the generators of SIDGAN \eg leveraging a mechanism for exploiting the temporal domain in the synthetic video data}
  
  \label{fig:pipeline}
\end{figure*}

\begin{figure*}[t!]
  \centering
  \includegraphics[scale=0.38]{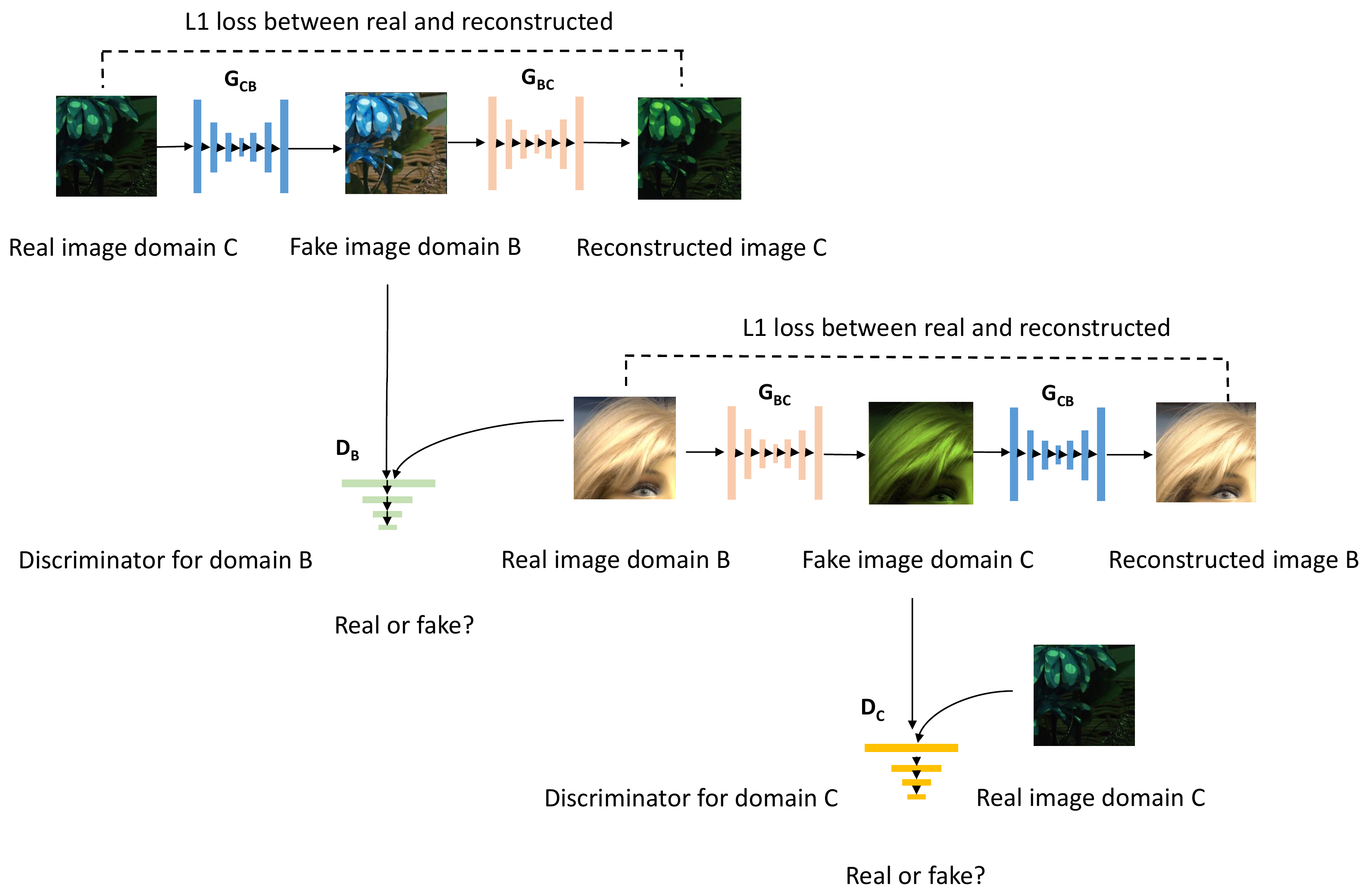}
  \caption{Visualisation of our Domain B-C CycleGAN, a sub-component of the complete dual CycleGAN architecture found in Figure~\ref{fig:pipeline}}

  \label{fig:domainsBC}
\end{figure*}

SIDGAN is modelled as a set of two CycleGANs in a dual configuration that learns the domain distributions for three domains; A, B and C. The model architecture is shown in Figure~\ref{fig:pipeline} and our domain B-C CycleGAN is shown in more detail in Figure~\ref{fig:domainsBC}. Domain A is characterised by a set of video frames defined by probability distribution $p_{A}$. The set of $N$ videos from this domain $\left\{V_{i}\right\}^{N}_{i=1} \sim p_{A}$ are available for training. Similarly, we also consider $M$ long exposure 
still images $\left\{L_{i}\right\}^{M}_{i=1} \sim p_{B}$ (domain B) and $T$ short exposure 
still images $\left\{S_{i}\right\}^{T}_{i=1} \sim p_{C}$ (domain C). The remainder of this section details how the sample sets are leveraged in order to learn a mapping from domain A to domain C via bridge domain B.

The A-B CycleGAN learns, in a conditional GAN fashion, an \emph{unpaired} mapping between domains A and B, transforming a set of RGB videos to a domain characterised by a set of RGB images (\ie long exposure images) using generators $G_{AB}$ and $G_{BA}$. This unsupervised CycleGAN does not require explicit sample pairings.  
Discriminator $D_{A}$ attempts to distinguish generated video frames $\hat{V}{=}G_{BA}(L)$, $L \sim p_{B}$ from real video frames drawn from the input distribution $V \sim p_{A}$ in Domain ${A}$ (Equation~\ref{eq:domainBA_gan}). Discriminator $D_{B}$ tries to differentiate synthetic long exposure still images $\hat{L}=G_{AB}(V)$, $V \sim p_{A}$ from real long exposure images $L \sim p_{B}$, drawn from Domain B (Equation~\ref{eq:domainAB_gan}).

\begin{equation}
\hspace{-1 em}   \mathcal{L}_{GAN}(G_{BA}, D_{A}) = \mathbb{E}_{L \sim p_B}  [\log(1 - \log(D_{A}(G_{BA}(L))))] +\mathbb{E}_{V_{\sim p_{A}}}  [\log(D_{A}(V))] 
    \label{eq:domainBA_gan}
\end{equation}

\begin{equation}
\hspace{-1 em}  \mathcal{L}_{GAN}(G_{AB}, D_{B}) = \mathbb{E}_{V \sim p_A} [\log(1 - \log(D_{B}(G_{AB}(V))))] +\mathbb{E}_{L_{\sim p_{B}}} [\log(D_{B}(L))] 
    \label{eq:domainAB_gan}
\end{equation}

We regularise the mappings between domains such that $G_{AB}$, $G_{BA}$ are approximate inverses of one another by employing a cycle consistency loss~\cite{zhu2017unpaired} (Equation~\ref{eq:cycle_loss}):

\begin{align}
\mathcal{L}_{cyc}(G_{AB}, G_{BA}) &= \mathbb{E}_{L \sim p_B} \|[  G_{AB}(G_{BA}(L)) - L]\|_{1} + \label{eq:cycle_loss} \\
       & +  \mathbb{E}_{V \sim p_A} \|[ G_{BA}(G_{AB}(V)) - V]\|_{1}.   \nonumber 
\end{align}

Following~\cite{TaigmanPW17,zhu2017unpaired}, we find it also important to add an identity loss $\mathcal{L}_{identity}$ in order to prevent colour inversion:

\begin{equation}
        \mathcal{L}_{identity}(G_{AB}, G_{BA}) = \mathbb{E}_{V \sim p_A} \|[ G_{BA}(V) - V]\|_{1}  +  \mathbb{E}_{L \sim p_B} \|[ G_{AB}(L) - L]\|_{1}.
    \label{eq:id_loss}
\end{equation}

Our final loss combines the introduced individual loss terms as a weighted combination, with individual components weighted by hyperparameters $\lambda_1$,$\lambda_2$ (Equation~\ref{eq:full}):

\begin{align}
\mathcal{L}(G_{AB},G_{BA}, D_{A}, D_{B}) &= \mathcal{L}_{GAN}(G_{BA}, D_{A}) + \mathcal{L}_{GAN}(G_{AB}, D_{B}) \label{eq:full} \\
       & + \lambda_1\mathcal{L}_{cyc}(G_{AB}, G_{BA}) + \lambda_2\mathcal{L}_{identity}(G_{AB}, G_{BA}).   \nonumber 
\end{align}

\begin{figure*}[t!]
\begin{center}
\begin{tabular}{c@{}c@{}c@{}}
    \scalebox{1}{Domain A} &
    \scalebox{1}{Domain B} &
    \scalebox{1}{Domain C} \\
\includegraphics[scale=0.46]{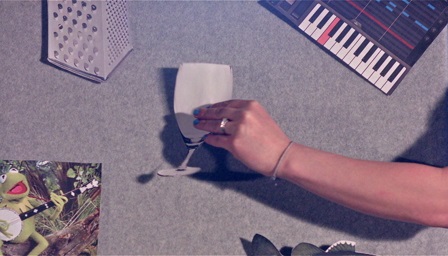} &
\includegraphics[scale=0.46]{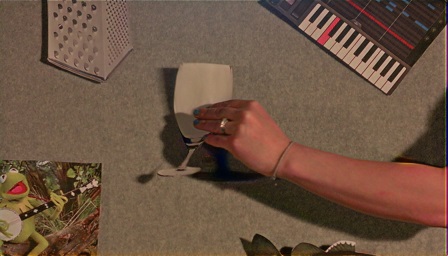} &
\includegraphics[scale=0.46]{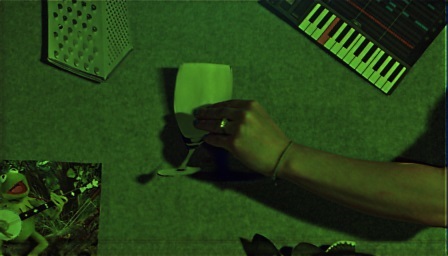} \\

\end{tabular}
\label{fig:synthetic_data_examples}
\caption{Generating synthetic long and short exposure frame pairs in two steps: \textbf{Step 1}: Project videos from domain A to our sensor-specific long exposure domain B using generator $G_{AB}$. \textbf{Step 2:} Project the translated image from step 1 to the sensor-specific short exposure domain C using generator $G_{BC}$ }
\end{center}
\end{figure*}

In contrast to the domain A-B mapping, the B-C CycleGAN is \emph{paired} (supervised) and employs generators $G_{BC}$ and $G_{CB}$. This component of our dual CycleGAN model is responsible for mapping long exposure RGB images to short exposure counterparts. This domain mapping is paired as, in contrast to dynamic video, it is easier to collect short-long exposure pairs for still images by using a tripod and varying camera exposure time. SIDGAN leverages this supervision, using intermediate domain B, with the aim of enhancing the quality of the target task; mapping dynamic videos (domain A) to short exposure (domain C). The B-C CycleGAN component employs a loss (Equation~\ref{eq:supervised1}), analogous to that of the domain A-B mapping, and additionally incorporates a $\mathcal{L}_{sup}$ term (Equation~\ref{eq:supervised2}), harnessing the supervisory signal that is available. 

\begin{align}
\mathcal{L}(G_{BC},G_{CB}, D_{B}, D_{C}) &= \mathcal{L}_{GAN}(G_{CB}, D_{B}) + \mathcal{L}_{GAN}(G_{BC}, D_{C}) \label{eq:supervised1} \\
       & + \lambda_{1}\mathcal{L}_{cyc}(G_{BC}, G_{CB}) + \lambda_{2}\mathcal{L}_{sup}(G_{BC}, G_{CB})  \nonumber 
\end{align}

Given a set of $M$ short-long exposure pairs $\left\{(S_{i}, L_{i})\right\}^{M}_{i=1}$, the supervised term $\mathcal{L}_{sup}(G_{BC}, G_{CB})$ is defined as:

\begin{equation}
\mathcal{L}_{sup}(G_{BC}, G_{CB}) = \mathbb{E}_{L \sim p_B} \|[ G_{BC}(L) - S]\|_{1}  +  \mathbb{E}_{S \sim p_C} \|[ G_{CB}(S) - L]\|_{1}.
\label{eq:supervised2}
\end{equation}

Our experimental evaluation (Section~\ref{sec:experiments}), demonstrates that a significant boost in translation quality is achieved when leveraging intermediate domain B to aid the weakly supervised CycleGAN mapping.

\subsection{Training Low-Light RAW-to-RGB Forward Models}
\label{sec:forward}

Our forward model training, fine-tuning schemes leverage a mixture of real and synthetic video data to learn a short-to-long exposure video mapping. We aim to extract an understanding of the correct colour and luminance distribution from real data (static video) while learning temporal consistency from the synthetic data (dynamic video). Our approach is shown in 
Figure~\ref{fig:pipeline}. In the first step, synthetic data is generated by taking internet videos and passing them through generators $G_{AB}$ and $G_{BC}$; using the process described previously (Section~\ref{sec:cyclegan}). In the second, step this synthetic video data is mixed with real data to train the forward model, adhering to the following three training and fine tuning steps:

\begin{enumerate}

\item \textbf{Training}: Train a 
forward model solely on real static video data 

\item \textbf{Fine Tuning \emph{a}}: Fine tune solely on synthetic dynamic video data 

\item \textbf{Fine Tuning \emph{b}}: Fine tune on real static video data 

\end{enumerate}

Our following experimental work (Section~\ref{sec:experiments}), employs a RAW-to-RGB forward model that follows the architecture of the SID motion model~\cite{Chen_2019_ICCV}, reproduced in Figure~\ref{fig:pipeline}. However, we note that our previously introduced synthetic data generation process is agnostic to specific model architectures. The model samples two frames from a static video and has three $L_{1}$ loss terms acting on the VGG features~\cite{SimonyanZ14a} of the two predicted frames ($\mathcal{L}_{a}$) and the two predicted frames and the groundtruth long exposure frame ($\mathcal{L}_{b}$, $\mathcal{L}_{c}$). As the training data is a static video there is no object and subject motion between frames, with noise being the only differentiator. We comment that while using our generator $G_{CB}$ to model the short-to-long exposure mapping would also be possible, we instead leverage a temporal consistency term in the forward model to exploit the temporal dimension of the synthetic video generated using the dual CycleGAN (Section~\ref{sec:cyclegan}).

\subsection{GAN Architectures for data generation}
\label{sec:arch}

Generators $G_{AB}$, $G_{BA}$, $G_{BC}$, $G_{CB}$ are modelled on the popular U-Net architecture~\cite{Ronneberger15}. In comparison to alternatives \eg ResNet, we corroborate previous work~\cite{Chen2018learning} and find that the encoder-decoder architecture of the U-Net to be amenable to high-quality image translation. Nevertheless, we modify the components of the U-Net to further increase the quality of the produced images. Our final generators are comprised of 5 convolutional blocks with a stride of 1  followed by $2{\times}2$ max pooling layers. Upsampling is performed using a nearest neighbour bilinear interpolation followed by a $1{\times1}$ convolution, which we found important to reduce the prevalence of checkerboard (upsampling) artifacts. 

The discriminators $D_{A}$, $D_{B}$, $D_{C}$ are all PatchGAN discriminators~\cite{pix2pix2016} which attempt to penalize structure at the scale of patches by classifying them as real or fake.  The discriminators ingest $192{\times}192$ patches which correspond to a receptive field that covers $75\%$ of the input image. Finally, we note that CycleGANs in our dual CycleGAN setup are optimised independently. Joint training is theoretically possible but poses a more difficult optimisation problem and exhausted our available GPU memory in practice. 

\section{Experimental Results}
\label{sec:experiments}

\subsection{Datasets and implementation details}

We employed the Vimeo-90K dataset~\cite{DBLP:journals/corr/abs-1711-09078} to translate real-world videos into our low-light sensor specific domain. The dataset has $91,701$ septuplet samples, each containing $7$ video frames of resolution $448{\times}255$. For the sensor-specific long and short exposure domains (\ie domains B and C), we use the Dark Raw Video (DRV) dataset~\cite{Chen_2019_ICCV}, which contains $224$ low-light raw video data and corresponding long-exposure images. Our intermediate domain B is represented by the long exposure DRV RGB images while for domain C we use the provided preprocessed DRV short exposure RAW video frames. 

Data is pre-processed using the pipeline of Chen~\etal\cite{Chen_2019_ICCV} which involves RAW-to-RGB conversion by averaging green pixels in each two-by-two block, black level subtraction, $2{\times}2$ binning, and global digital gain. 
Furthermore, noise is reduced using VBM4D~\cite{Maggioni2012VideoDD} and pixel values are linearly scaled using exposure value (EV) difference. We resize the DRV long and short exposure RGB images such that 
resolution matches that of Vimeo-90K and normalize images in $[-1,1]$. Experimentally we find that training on large patches is crucial in order to capture the global statistics and learn the correct white balance. For this reason, both CycleGAN$_{AB}$ and CycleGAN$_{BC}$ are trained using $256{\times}256$ crops corresponding to $50\%$ of the resized DRV frames. We randomly select $400$ Vimeo-90K videos and train our Dual CycleGAN, retaining the original train/val/test partitions of the DRV dataset. 
Finally, our forward RAW-to-RGB model is trained using the train partition of the DRV dataset and $9,366$ synthetic videos.  

Models are implemented using Tensorflow and Keras \cite{keras,DBLP:journals/corr/AbadiABBCCCDDDG16} and trained using an NVidia Tesla V100 GPU with 32GB memory. Our CycleGAN models are trained 
initially for $50$ epochs with a learning rate $10^{-4}$ which then linearly decays for a further $20$ epochs. 
Hyperparameters $\lambda_{1}$, $\lambda_{2}$ are found by empirical search and set to values 6.0, 6.0 in Equation~\ref{eq:full}, and values 10.0, 10.0 in Equation~\ref{eq:supervised1}, respectively. 
The batch size is set to 1 and our forward model is trained using the training scheme described in Section~\ref{sec:forward} for a total of $1000$ epochs. We employ a learning rate of $10^{-4}$ for the initial $500$ epochs and reduce this to $10^{-5}$ for the latter half of training. 

\subsection{Synthetic data quality evaluation} 

\begin{figure*}[t!]
\begin{center}
\begin{tabular}{c@{}c@{}c@{}c@{}}
    \scalebox{0.68}{(KID 5.06)} &
    \scalebox{0.68}{(KID 4.78)} & 
    \scalebox{0.68}{ (KID 4.68)} & 
    \scalebox{0.68}{(KID 3.99)} \\
\includegraphics[scale=0.38]{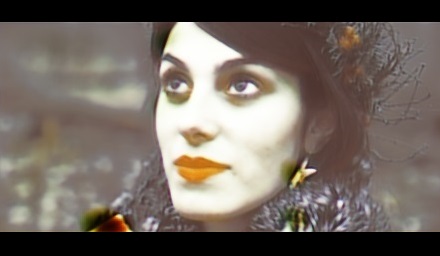} &
\includegraphics[scale=0.38]{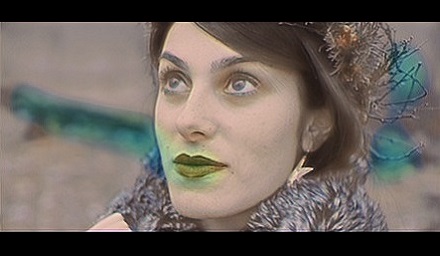} &
\includegraphics[scale=0.38]{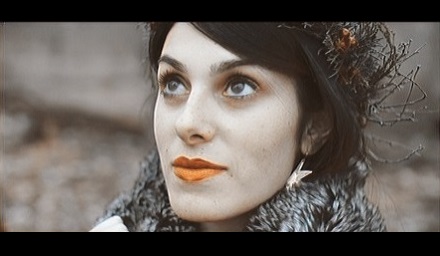} &
\includegraphics[scale=0.38]{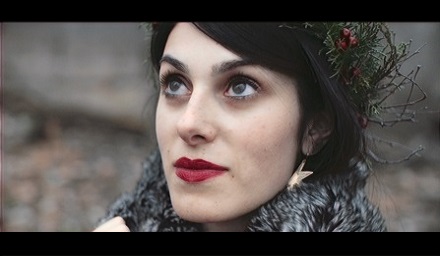} 
\end{tabular}
\end{center}
\caption{Evolution of the KID distance during the unpaired training of CycleGAN$_{AB}$. The KID distance correlates well with visual quality and is used for model selection}
\label{fig:kid_comparison}
\end{figure*}

We distinguish between the unpaired task that pertains to CycleGAN$_{AB}$ and the paired RGB-to-RAW mapping of CycleGAN$_{BC}$. Since CycleGAN$_{AB}$ is responsible for mapping videos from any source to our sensor-specific domain, no ground truth information is available for this task. In order to numerically evaluate generators $G_{AB}$ and $G_{BA}$ we adopt the following metrics: Fr\'echet Inception Distance (FID)~\cite{DBLP:journals/corr/HeuselRUNKH17} and Kernel Inception distance (KID)~\cite{DBLP:journals/corr/abs-1802-03446}. 
For CycleGAN$_{BC}$, we use the available ground truth; long and short exposure pairs of the test partition (DRV dataset) and evaluate performance using standard metrics; Peak Signal-to-Noise Ratio (PSNR), Structural Similarity (SSIM).

\begin{figure*}
\begin{center}
\begin{tabular}{c@{}c@{}c@{}c@{}}
    \scalebox{0.62}{Long exp. ground truth } &
    \scalebox{0.62}{Short exp. ground truth } &
    \scalebox{0.62}{Prediction (PSNR 32.9)} &
    \scalebox{0.62}{Error heatmap} \\
\includegraphics[scale=0.35]{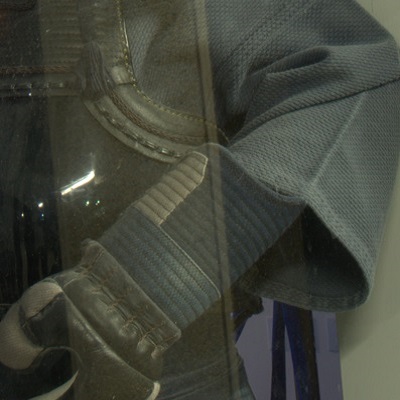} &
\includegraphics[scale=0.35]{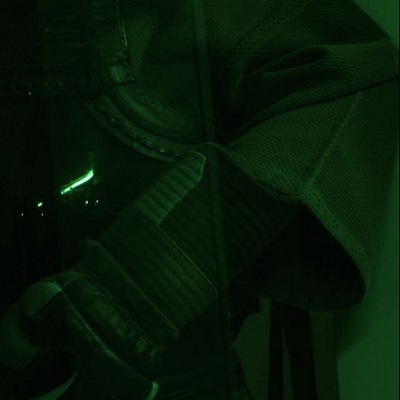} &
\includegraphics[scale=0.35]{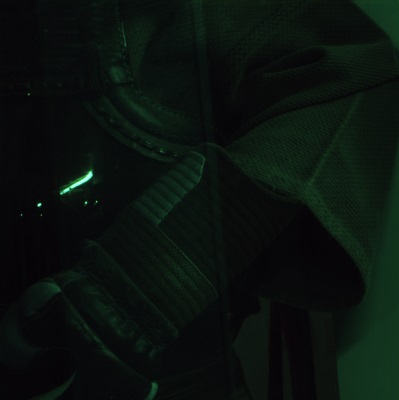} &
\includegraphics[scale=0.35]{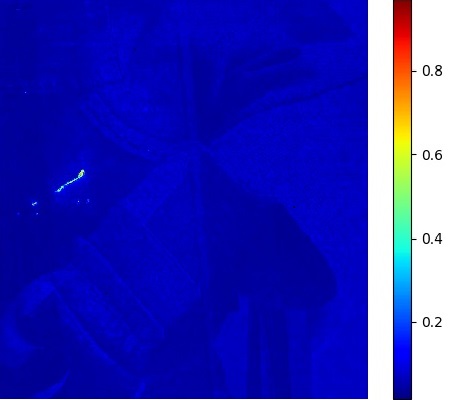} \\
\end{tabular}
\begin{tabular}{c@{}c@{}c@{}c@{}}
    \scalebox{0.60}{Long exp. ground truth } &
    \scalebox{0.60}{Short exp. ground truth } &
    \scalebox{0.60}{Prediction (PSNR 27.93)} &
    \scalebox{0.60}{Error heatmap} \\
\includegraphics[scale=0.35]{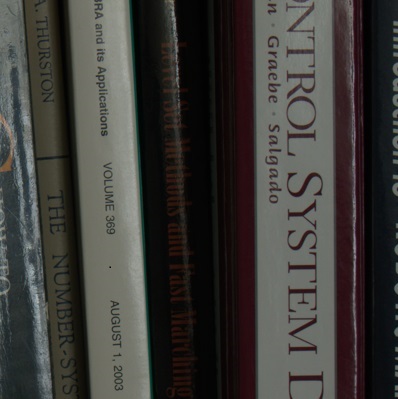} &
\includegraphics[scale=0.35]{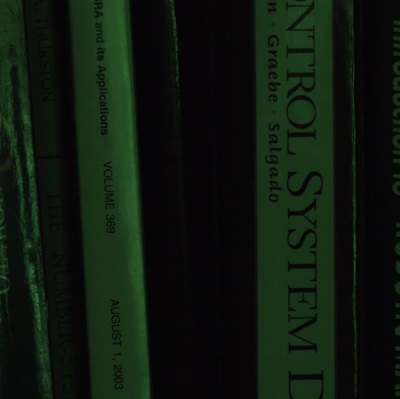} &
\includegraphics[scale=0.35]{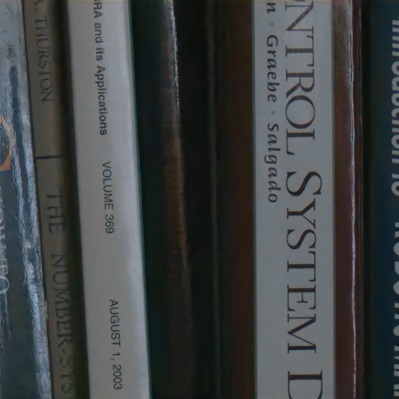} &
\includegraphics[scale=0.35]{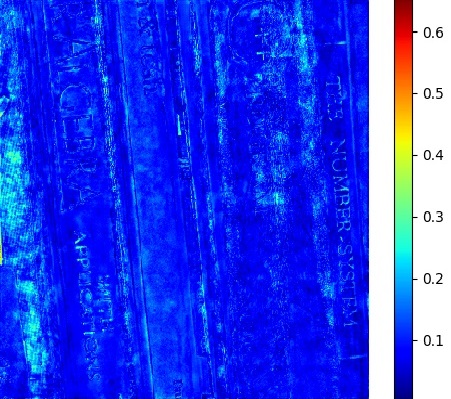} 
\end{tabular}
\end{center}
\caption{First row: Mapping from domain B (long exposure) to domain C (short exposure) using generator $G_{BC}$. Second row:  Mapping from domain C (short exposure) to domain B (long exposure) using generator $G_{CB}$} 
\label{fig:visual_comparison2}
\end{figure*}

We observe experimentally that the KID 
correlates better than FID with the visual quality of the generated samples (see Fig.~\ref{fig:kid_comparison}) and we base final selection of models $G_{BA}$ and $G_{AB}$ solely on KID score. 
Our generator $G_{BC}$, responsible for mapping long exposure (domain B) to short exposure (domain C), achieves $27.28 dB$ PSNR and $0.88$ SSIM and $G_{CB}$ performance is $25.28 dB$ and $0.74$, respectively. 
Quantitative results allude to the fact that long exposure captures more photons and images better represent scene colors and contrast. Intuitively the problem can be regarded as more ill-posed when mapping in the short to long direction. We also observe by ablation that training without the supervised term (Equation~\ref{eq:supervised2}) resulted in significantly lower performance (${\sim}4dB$ less), providing evidence in support of our choice to decompose the data synthesis task into two separate learning problems and exploit the available paired data via the intermediate domain mapping. Figure~\ref{fig:visual_comparison2} provides example predictions for generators $G_{CB}$ and $G_{BC}$.
We compare our trained RAW-to-RGB forward model against state-of-the-art approaches for low-light image and video processing. Following~\cite{Jiang_2019_ICCV}, we evaluate the performance on the static videos of the DRV dataset and examine both the image quality and the temporal stability of our method.

\begin{figure*}[t!]
\begin{center}
\begin{tabular}{c@{}c@{}c@{}}
    \scalebox{0.68}{Ground Truth} &
      \scalebox{0.68}{Unsup. (PSNR 24.0)} & 
      \scalebox{0.68}{Semi-sup. (PSNR 27.2)} \\
     \includegraphics[scale=0.52]{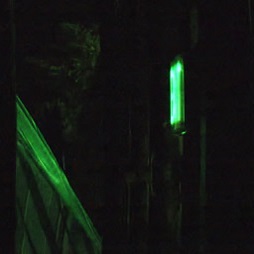} &
\includegraphics[scale=0.52]{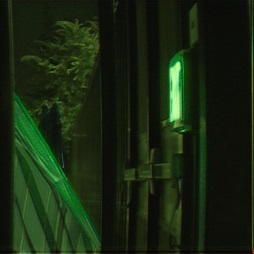} &
\includegraphics[scale=0.52]{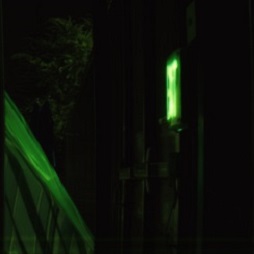} 
\end{tabular}
\end{center}
\caption{Comparing CycleGAN with the proposed semi-supervised CycleGAN. Our semi-supervised variant shows better translation performance by exploiting the ground truth information in the optimization objective}
\label{fig:unsupervised_comparison}
\end{figure*}

\begin{figure*}[t!]
\begin{center}
\begin{tabular}{c@{}c@{}c@{}}
    \scalebox{1}{Ground Truth} &
      \scalebox{1}{SID motion} & 
      \scalebox{1}{Ours (SIDGAN)} \\
     \includegraphics[scale=0.077]{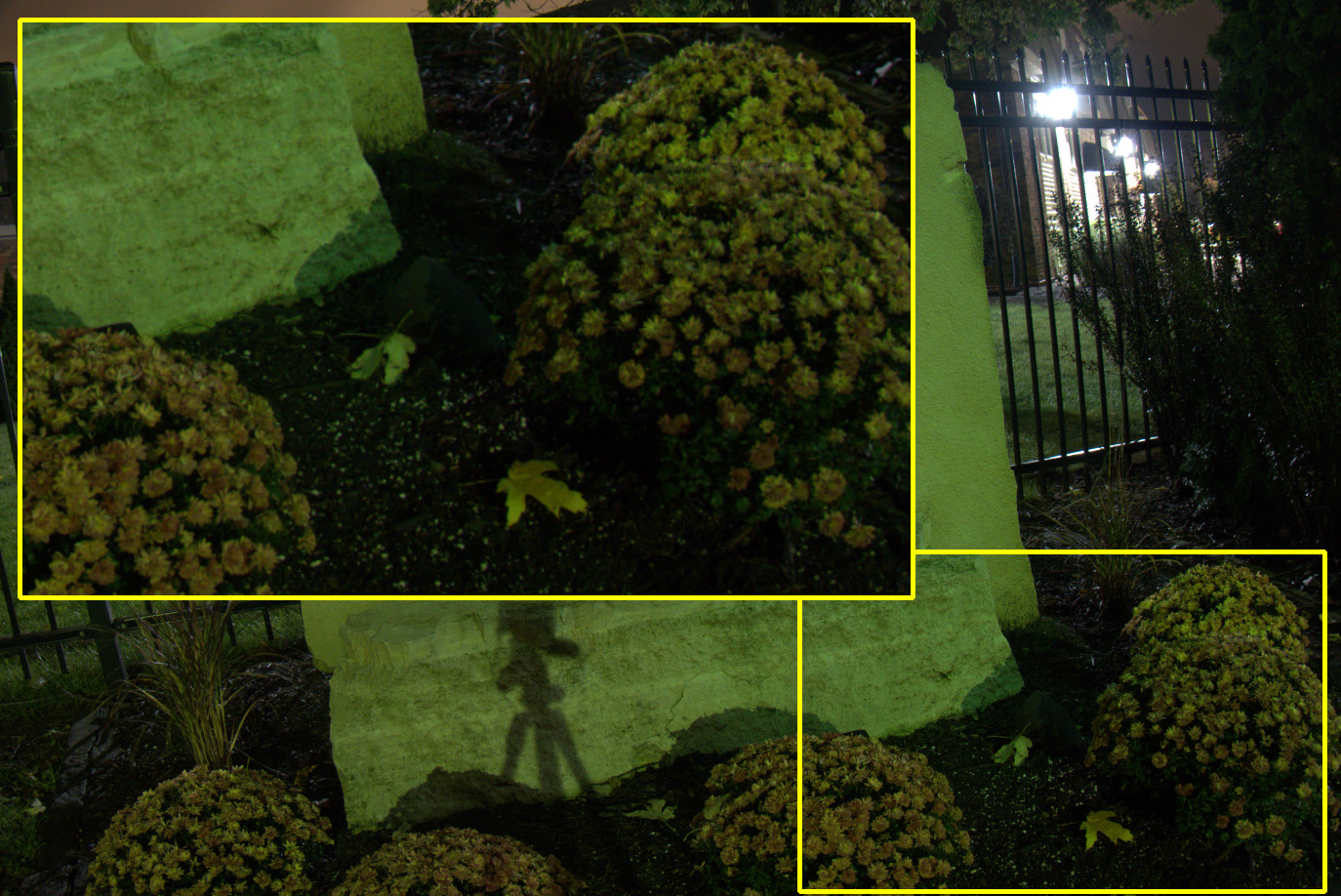} &
\includegraphics[scale=0.077]{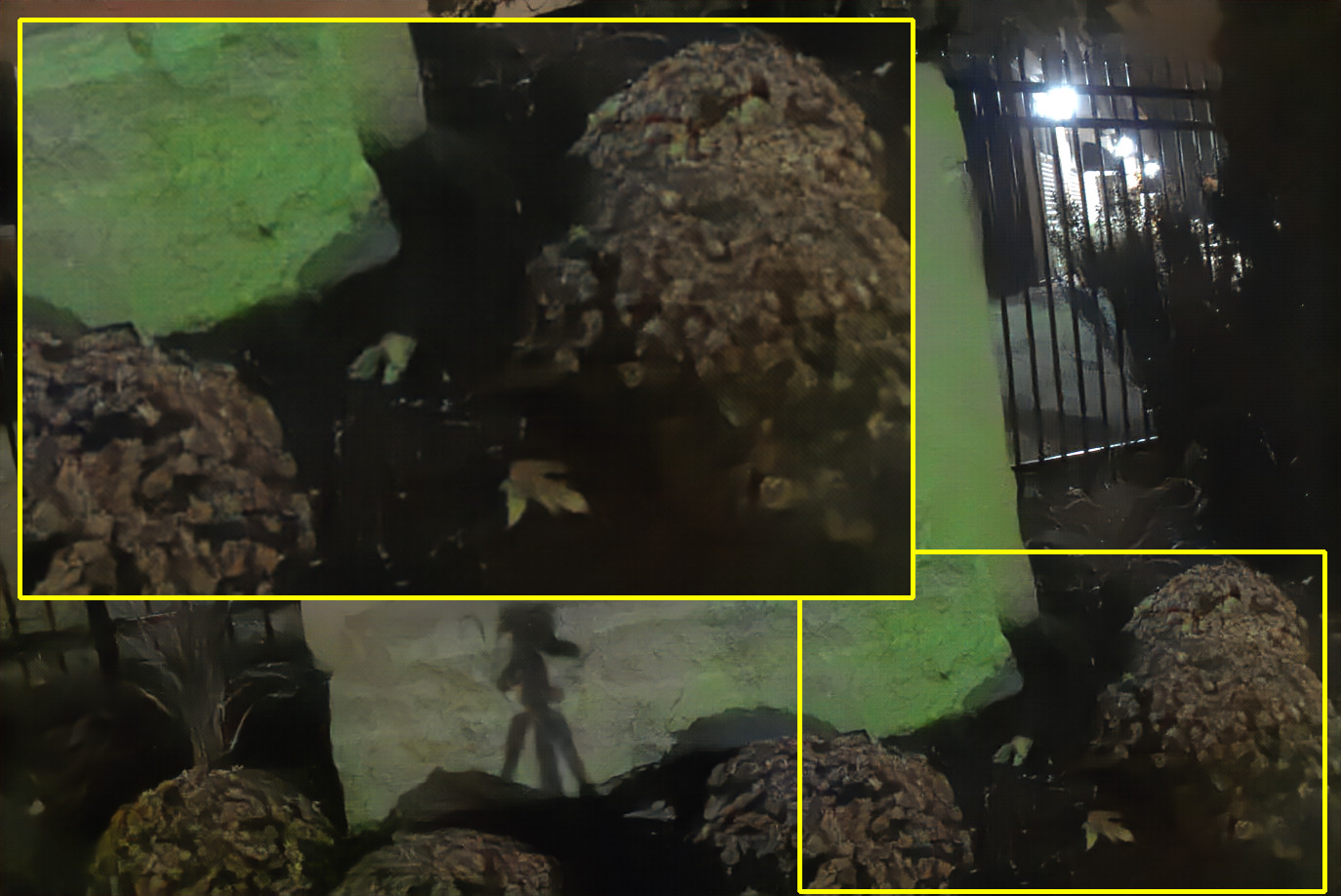} &
\includegraphics[scale=0.077]{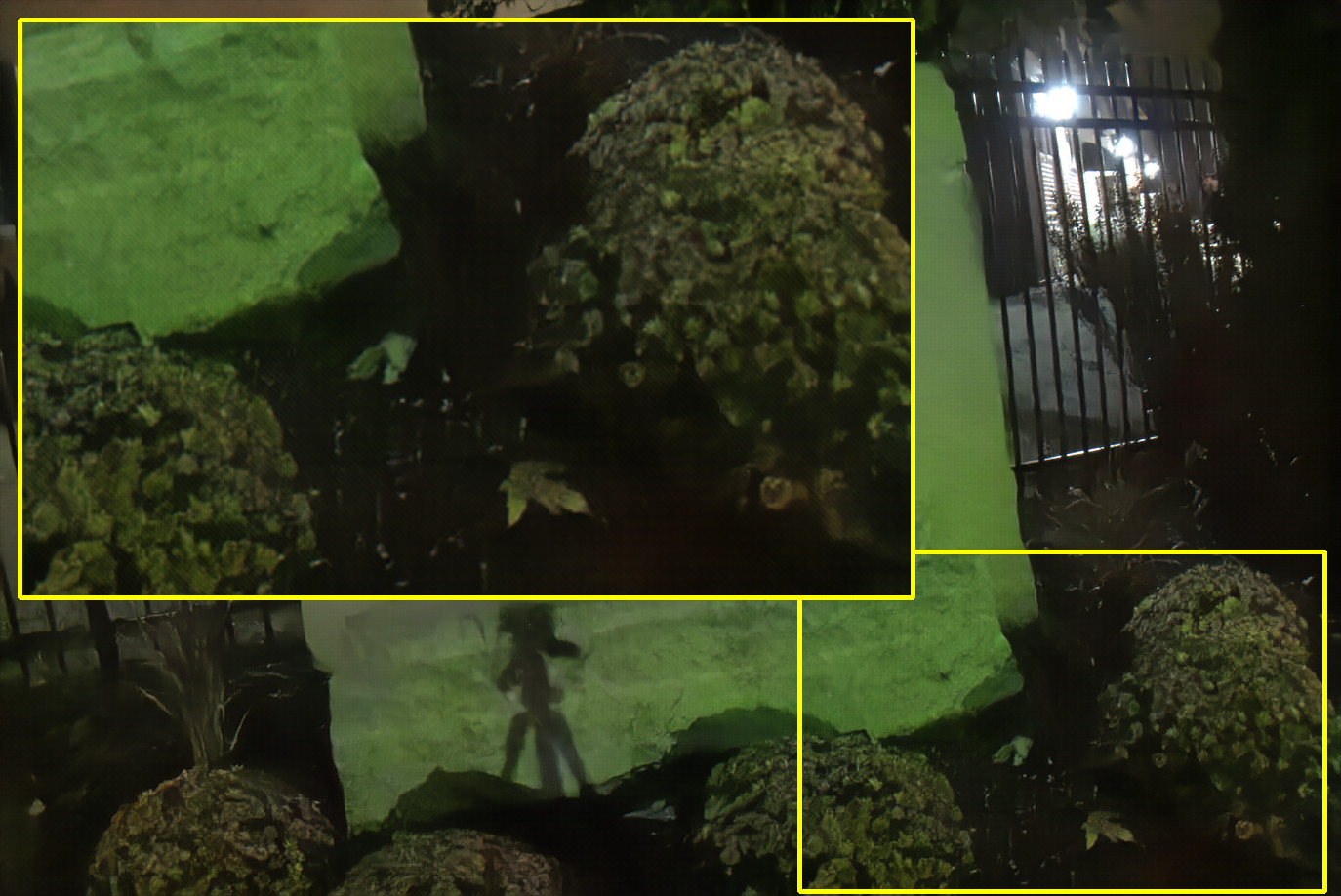} 
\end{tabular}
 \begin{tabular}{c@{}c@{}c@{}}
      \includegraphics[scale=0.077]{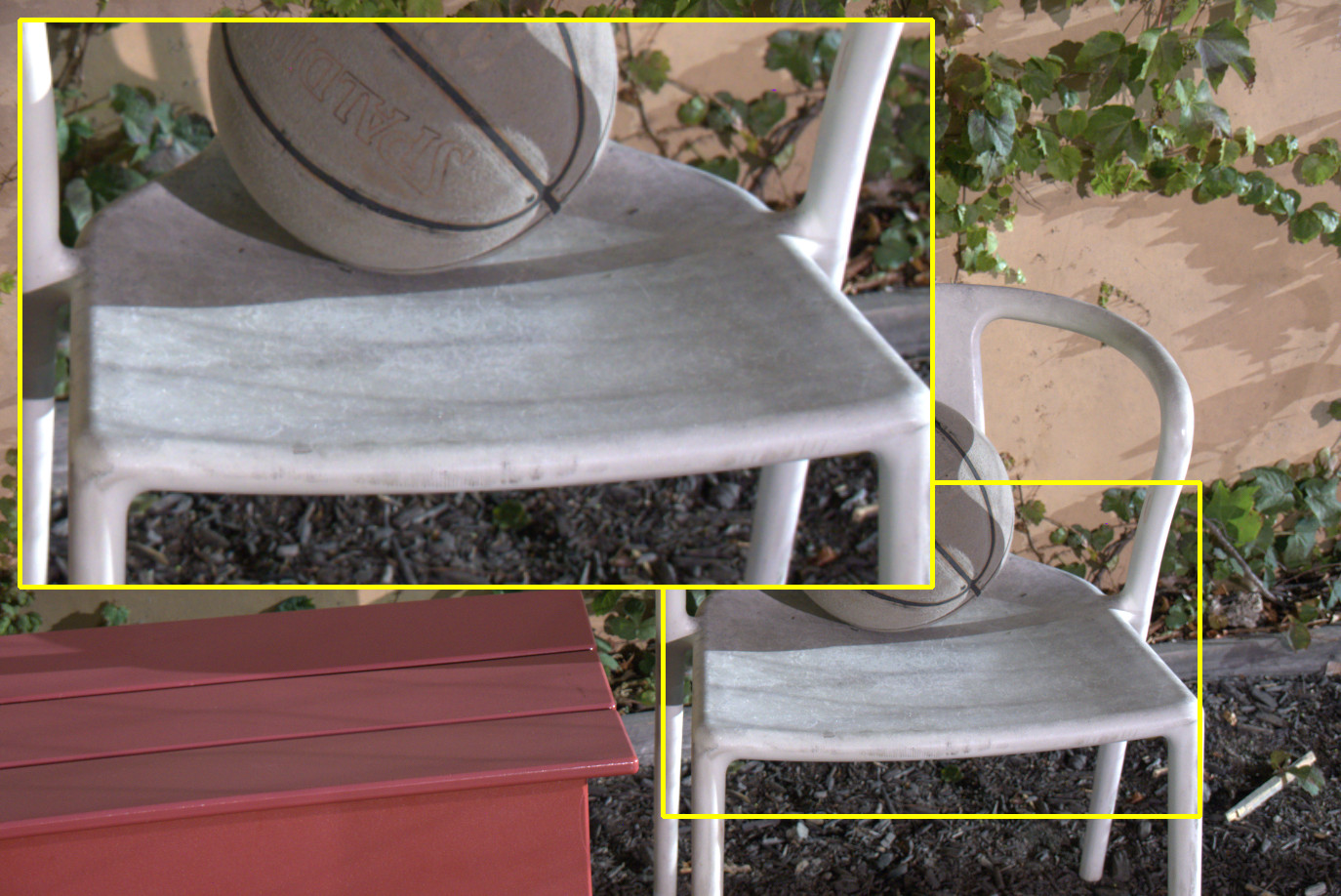} &
 \includegraphics[scale=0.077]{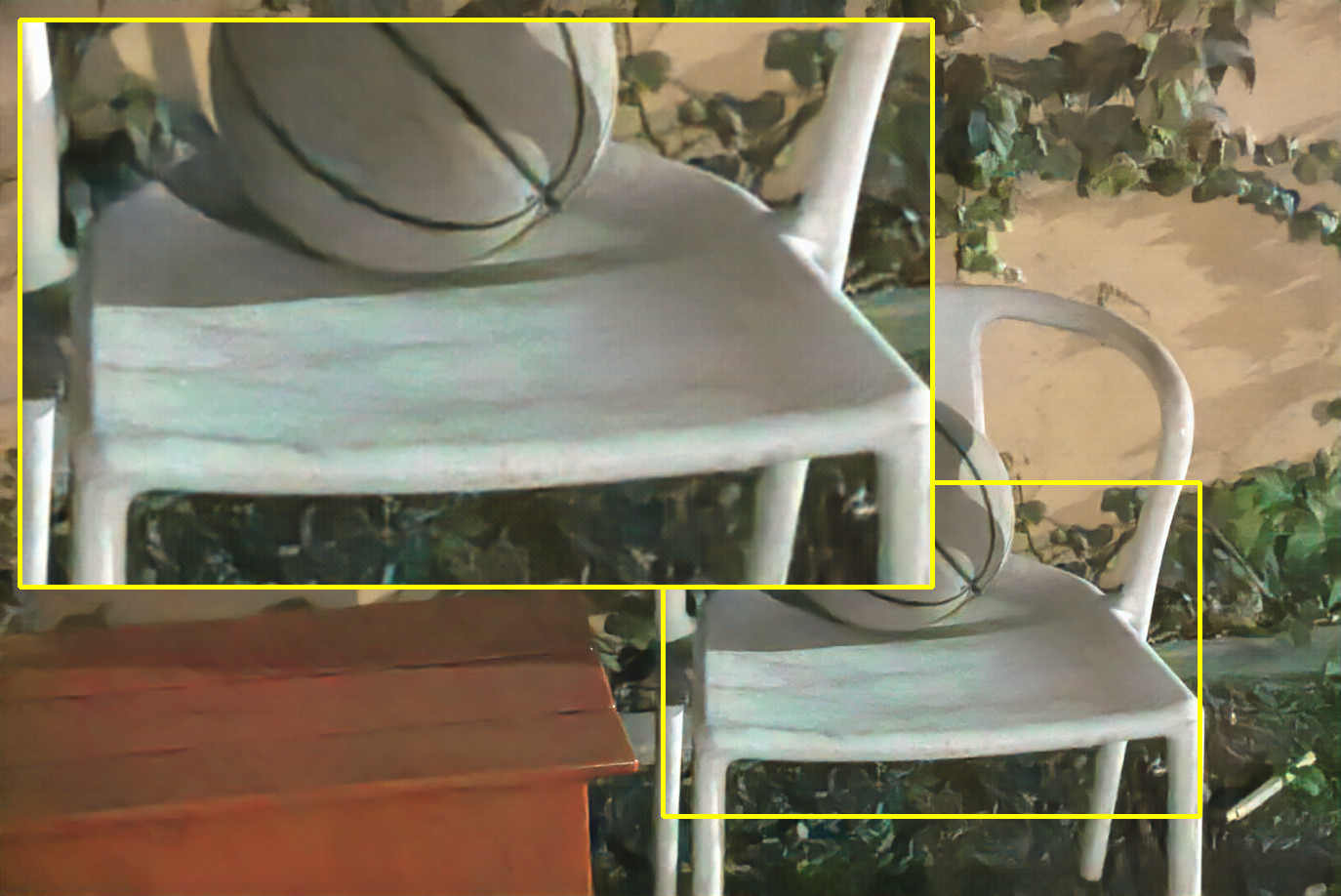} &
 \includegraphics[scale=0.077]{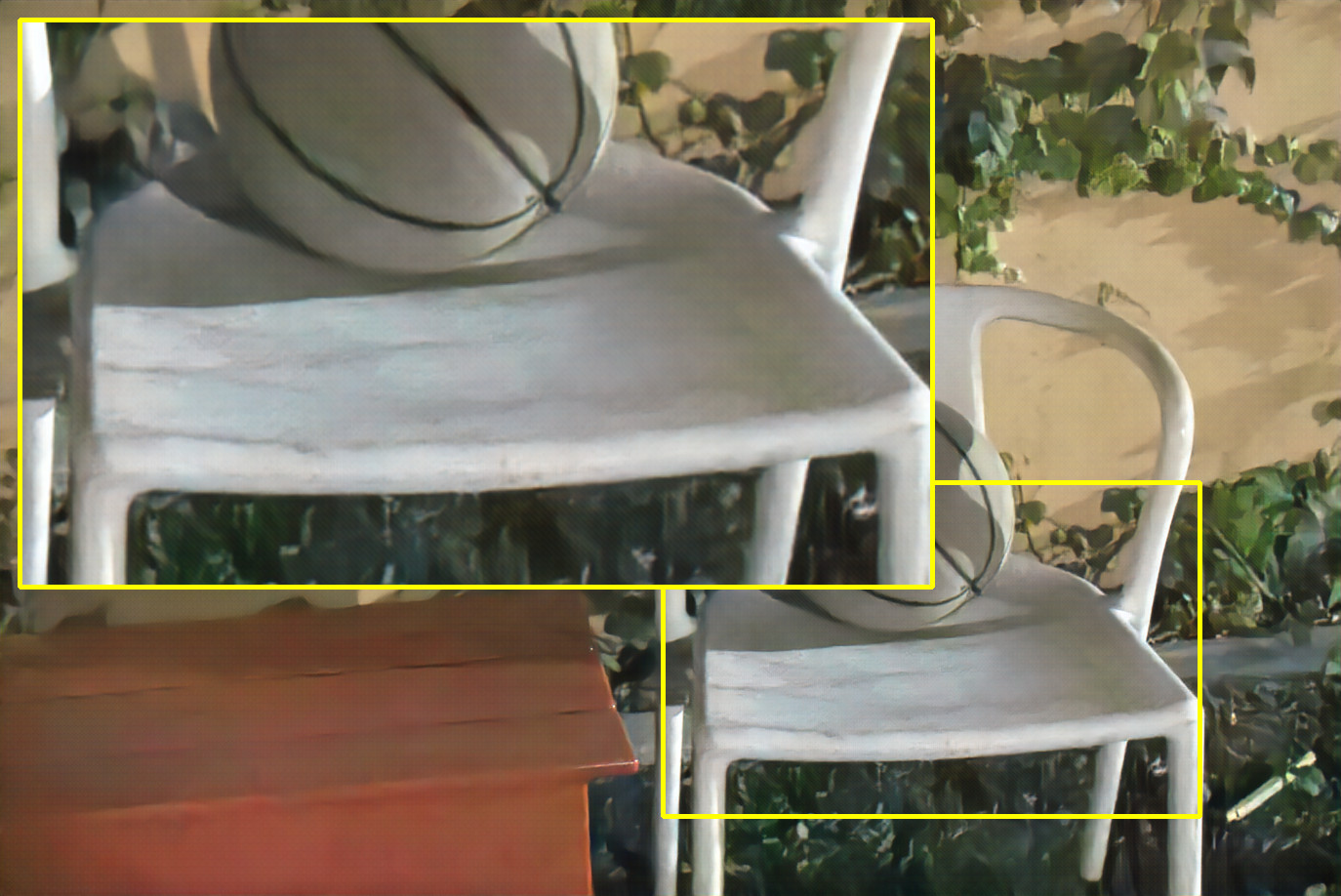} 
 \end{tabular}

\end{center}

\caption{Comparing the image quality with SID motion \cite{Chen_2019_ICCV}. Note the improved colours in the marked regions}

\label{fig:visual_comparison_forward}
\end{figure*}
\begin{figure}[]
\centering
\begin{tabular}{c@{}c@{}}
\includegraphics[scale=0.25]{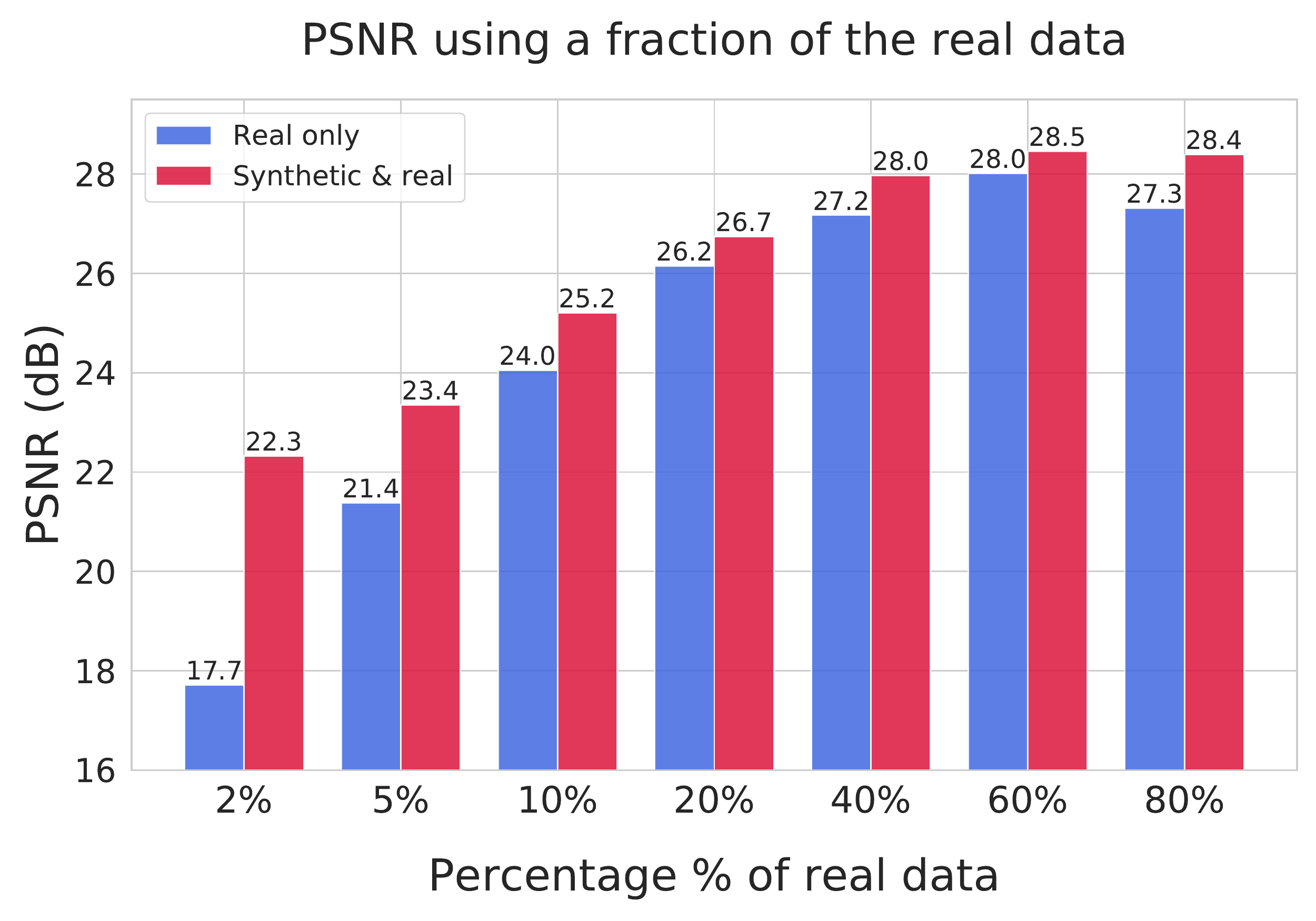} &
\includegraphics[scale=0.25]{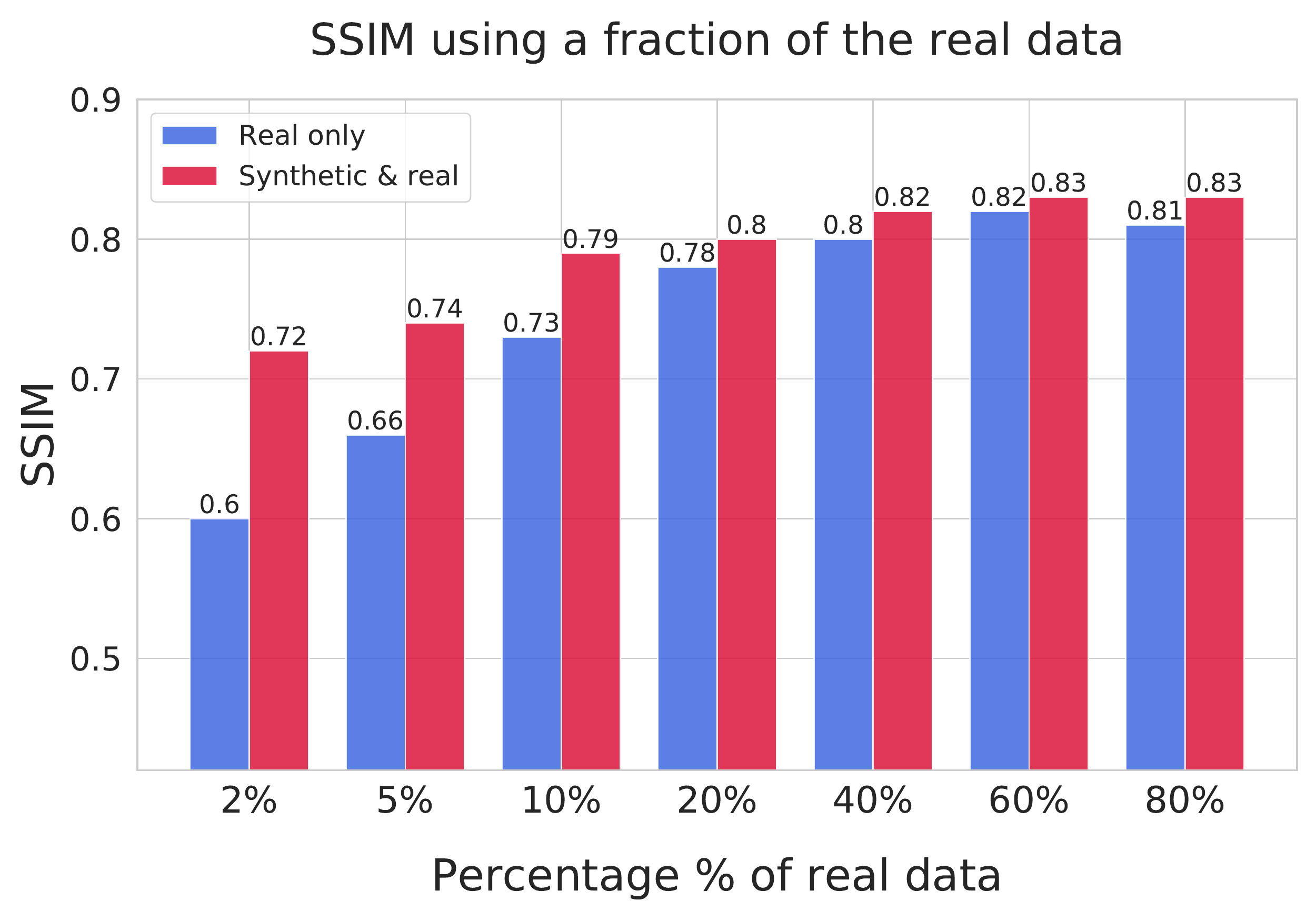} 
\end{tabular}
\caption{The effect of real : synthetic training data ratios using portions of the DRV dataset. Left: PSNR, right: SSIM}
\label{fig:ablation_1}
\end{figure}
\subsection{Output image and video quality evaluation}
\label{sec:img_and_vid_qual_eval}

\begin{table}[ht]
 \setlength\tabcolsep{7pt}
 \def\arraystretch{1.1}
\centering
\caption{Output image quality on the DRV static dataset} 
 \begin{tabular}{ p{40mm} | p{20mm} p{20mm} }
 \toprule
\textbf{Model} & \textbf{PSNR}$\uparrow$& \textbf{SSIM}$\uparrow$  \\
\hline
Input+Rawpy            & ${12.94}$ & ${0.165}$  \\ 
VBM4D+Rawpy            & ${14.77}$ & ${0.315}$  \\ 
KPN+Rawpy              & ${18.81}$ & ${0.540}$  \\ 
SID w/o VBM4D          & ${27.32}$ & ${0.799}$  \\ 
SID                    & ${27.69}$ & ${0.803}$  \\ 
SID Motion (real only) & ${28.26}$ & ${0.815}$  \\ 
\hline
\hline
SIDGAN (synthetic only)   & $21.53$ & $0.704$  \\ 
SIDGAN (synthetic + real) & \textbf{28.94} & \textbf{0.830} \\
\bottomrule
\end{tabular}
\label{tab:image_quality}
\end{table}

\begin{table}[ht]
\setlength\tabcolsep{7pt}
\def\arraystretch{1.1}
\centering
\caption{Output video quality on the DRV static dataset}
\label{tab:video_quality}
\begin{tabular}{ p{40mm} |  c c}
\toprule

\textbf{Model} & \textbf{TPSNR}$\uparrow$ & \textbf{TSSIM}$\uparrow$  \\
\hline
SID [9] w/o VBM4D    & $33.72$ & $0.939$   \\ 
SID  & ${37.05}$ & ${0.961}$    \\ 
SID Motion (real only)  & ${38.31}$ & $\textbf{0.974}$    \\ 
\hline
\hline
SIDGAN (synthetic + real )& \textbf{39.34} & 0.966  \\ 
\bottomrule
\end{tabular}
\end{table}
 
\begin{table}
\setlength\tabcolsep{7pt}
\def\arraystretch{1.1}
\centering
\caption{Dynamic video quality evaluation with varying real data ratios. Training the same model using increasing fractions of real data only (SID motion) and real data and synthetic data (SIDGAN).}
\label{tab:temporal_quality}
\begin{tabular}{ p{65mm} |  c }
\toprule
\textbf{Model} & $E_{warp} {\times}10^{-5} \downarrow$ \\
\hline
SID Motion (2\% real DRV data)   & 55.9  \\ 
SID Motion (5\% real DRV data)   & 54.3  \\ 
SID Motion (20\% real DRV data)  & 35.6  \\ 
SID Motion (100\% real DRV data) & 29.3  \\ 
\hline
\hline
SIDGAN (2\% real DRV data + synthetic)  & 31.2          \\ 
SIDGAN (5\% real DRV data + synthetic)  & 32.7          \\ 
SIDGAN (20\% real DRV data + synthetic) & 32.2          \\ 
SIDGAN (100\% real DRV data + synthetic)& \textbf{28.2} \\ \bottomrule
\end{tabular}
\end{table}

\noindent{\textbf{Image Quality:}} for consistent comparison with previous work~\cite{Chen_2019_ICCV}, we compare the fifth frame of our output video with the respective long exposure ground truth image and evaluate the performance in terms of average PSNR and SSIM over the $49$ DRV test videos. We compare performance with recent methods SID~\cite{Chen2018learning} and SID motion~\cite{Chen_2019_ICCV} as well as common baselines that combine performant denoising algorithms (VBM4D~\cite{Maggioni2012VideoDD}, KPN~\cite{DBLP:journals/corr/abs-1712-02327}) with traditional non-learning based enhancement tools (here using Rawpy\footnote{\url{https://pypi.org/project/rawpy/}}). Results are summarised in Table~\ref{tab:image_quality}. Baselines are observed to perform poorly for this challenging task. Our forward model, trained purely on synthetic data, achieves a PSNR of $21.53dB$ and SSIM of $0.70$. We attribute this fairly weak performance to a well understood domain shift between synthetic and real data~\cite{nowruzi2019much,sankaranarayanan2018learning}. %
However, we observe that training the model by adding a small fraction of real data (with a real : synthetic data ratio of $1\!:\!45$) successfully diminishes this domain gap yielding $28.94dB$ PSNR and $0.83$ SSIM, constituting state-of-the-art performance on the DRV dataset.

\noindent{\textbf{Temporal Consistency:}} the DRV dataset contains static raw videos, thus temporal stability can be measured by computing temporal PSNR (TPSNR) and temporal SSIM (TSSIM) between pairs of consecutive frames, in similar fashion to~\cite{Chen2018learning}. Results are presented in Table~\ref{tab:video_quality}. Our model offers competitive results when evaluated under these temporal metrics and we attribute strong performance to the extra information provided by our dynamic video synthetic data. We further evaluate dynamic video temporal stability by introducing synthetic training data, in a varying ratio with (scarce) real data. Average temporal warping error~\cite{Lai-ECCV-2018} is reported in Table~\ref{tab:temporal_quality}. Largest improvements are observed when available real data is scarcest. 
\subsection{Real training data quantity and ratios}

The addition of real image data was shown to help close synthetic training distribution domain gaps, resulting in quantitative improvements (Section~\ref{sec:img_and_vid_qual_eval}). We further investigate the effect of adding real image data quantities in relation to synthetic data. Subsets of the DRV real dataset comprising 2\%, 5\%, 10\%, 20\% 40\%, 60\% and 80\% are randomly sampled and model performance is evaluated when training solely on these real data subsets. We additionally train models on a set of $9,366$ synthetic videos, generated by SIDGAN, and then fine-tune with the aforementioned real data subsets accordingly. All models are trained for $1000$ epochs using identical hyperparameters. PSNR and SSIM performance is reported in Figure~\ref{fig:ablation_1}. We observe that the addition of our synthetic data significantly boosts performance; increasing PSNR from $17.70$ to $22.32$, from $21.35$ to $23.35$ and from $24.04$ to $25.19$ for the cases of $2\%, 5\%$ and $10\%$, respectively. As the fraction of real data is increased, the gap in performance reduces indicating that the addition of our synthetic data again offers largest benefit in scenarios where real data is scarce.

\section{Conclusions}

We introduce \textbf{S}eeing \textbf{I}n the \textbf{D}ark \textbf{GAN} (SIDGAN), a data synthesis method addressing the training data bottleneck encountered when learning models for RAW-to-RGB problems. SIDGAN comprises two CycleGANs in order to leverage an intermediate domain mapping. Tasks that involve mapping between domains containing disparate appearance yet also lacking paired samples, can benefit from \emph{intermediate domain} mappings that possess a paired data relationship with one of the original domains. We show that this strategy is capable of increasing the strength of the training signal and results in significant improvements for the investigated low-light RAW-to-RGB problem. Such tools may be widely applicable for domain mapping instances where data collection of directly paired samples between the domains of interest proves difficult or impossible. 

\bibliographystyle{splncs04}
\bibliography{egbib}

\end{document}